\documentstyle[pre,aps,epsf,twocolumn]{revtex}
\newcommand{\beq}{\begin{equation}}
\newcommand{\eeq}{\end{equation}}
\newcommand{\bea}{\begin{eqnarray}}
\newcommand{\eea}{\end{eqnarray}}
\begin{document}
\draft
\title{Coulomb crystals in the harmonic lattice approximation}
\author{D.~A. Baiko and D.~G. Yakovlev}
\address{Ioffe Physical--Technical Institute,
     194021 St.--Petersburg, Russia}
\author{H.~E. De Witt}
\address{Lawrence Livermore National Laboratory, CA 94550 Livermore}
\author{W.~L. Slattery}
\address{Los Alamos National Laboratory, NM 87545 Los Alamos}

\date{\today}
\maketitle

\begin{abstract}
The dynamic structure factor
${\tilde S}({\bf k},\omega)$ and the
two-particle distribution
function $g({\bf r},t)$ of ions in a Coulomb crystal
are obtained in a closed analytic form using the harmonic
lattice (HL) approximation which takes into account
all processes of multi-phonon excitation and absorption.
The static radial two-particle distribution function $g(r)$
is calculated for classical
($T \gtrsim \hbar \omega_p$, where
$\omega_p$ is the ion plasma frequency) and
quantum ($T \ll \hbar \omega_p$)
body-centered cubic (bcc) crystals. The results for
the classical crystal are in a very good agreement with
extensive Monte Carlo (MC) calculations at $1.5 \lesssim r/a \lesssim 7$,
where $a$ is the ion-sphere radius.
The HL Coulomb energy is calculated for classical
and quantum bcc and face-centered cubic crystals,
and anharmonic corrections are discussed.
The inelastic part of the HL static structure
factor $S''(k)$, averaged over orientations of wave-vector {\bf k},
is shown to contain pronounced singularities
at Bragg diffraction positions. The type of the singularities is 
different in classical and quantum cases.
The HL method can serve as a useful tool complementary to
MC and other numerical methods.
\end{abstract}


% 50. PHYSICS OF GASES, PLASMAS, AND ELECTRIC DISCHARGES
% 52. Physics of plasmas and electric discharges
% 52.25.-b: Plasma properties
% 52.25.Kn: Thermodynamics of plasmas
% 52.25.Fi: Transport properties
% 52.25.Tx: Emission, absorption, and scattering of particles
% 52.25.Ub Strongly-coupled plasmas
% 52.25.Zb Dusty plasmas; plasma crystals
% 72. Electronic transport in condensed matter
% 72.10.-d Theory of electronic transport, scattering mechanisms
% 72.10.Di: Scattering by phonons, magnons, and other nonlocalized excitations
% 95. Fundamental A&A
% 95.30.Qd: MHD and plasmas
% 97.20.Rp: Faint blue stars, white dwarfs, degenerate stars, nuclei of
%planetary nebulae
% 97.60.Jd: Neutron stars

%%%%%%%%%%%%%%%%%%%%  GENERAL CONSIDERATION  %%%%%%%%%%%%%%%%%%%%%%
%%%%%%%%%%%%%%%%%%%%%%%%%% Sect. 1 %%%%%%%%%%%%%%%%%%%%%%%%%%%%%%%%
\section{Introduction}

A model of a Coulomb crystal of point charges in a uniform
neutralizing background of charges of opposite sign
is widely used in various branches of physics.
The model was originally proposed by Wigner \cite{Wigner34}
who showed that zero-temperature electron gas
immersed into uniform background of positive charges
crystallizes into body-centered cubic (bcc) Coulomb crystal
at sufficiently low density.
Since then the model has been used in solid state
physics for describing electron-hole plasma
(e.g., Ref.\ \cite{Rakhmanov78}) and
in plasma physics for describing dusty plasmas and
ion plasmas in Penning traps
(e.g., Ref.\ \cite{IBTJHW98}). Finally, Coulomb crystals
of ions on almost uniform background of degenerate electron gas
are known to be formed in the cores of white dwarfs
and the envelopes of neutron stars. Consequently, properties of
Coulomb crystals are important for studying structure
and evolution of these astrophysical objects
(e.g., Ref.\ \cite{C93}).

As classical examples of strongly coupled systems,
the Coulomb crystals have been the subject of
extensive studies by various numerical methods,
mostly by Monte Carlo (MC; e.g., \cite{SDS90},
and references therein), and also by molecular dynamics
(MD; e.g., Ref.\ \cite{FH93}), and path-integral Monte
Carlo (PIMC; e.g, Ref.\ \cite{O97}). Although the
results of these studies are very impressive,
the numerical methods are time consuming and
require the most powerful computers.

The aim of the present article is to draw attention
to a simple analytic model of Coulomb crystals.
It has been employed recently in Ref.\ \cite{BKPY98}
in connection with transport properties of degenerate
electrons in strongly coupled plasmas of ions.
We will show that this model is 
a useful tool for studying static and dynamic
properties of Coulomb crystals themselves.

%%%%%%%%%%%%%%%%%%%%%% Section 2 %%%%%%%%%%%%%%%%%%%%%%%%%%
\section{Structure factors in harmonic lattice approximation}

For certainty, consider a Coulomb crystal of ions immersed in
a uniform electron background.
Let $\hat{\rho}({\bf r},t)= \sum_i \delta({\bf r}-\hat{{\bf r}}_i(t))$
be the Heisenberg representation operator of the ion number density, where
$\hat{{\bf r}}_i(t)$ is the operator of the $i$th ion position.
The spatial Fourier
harmonics of the number density operator is
$\hat{\rho}_{\bf k}(t)=\sum_i e^{-\imath{\bf k}\cdot\hat{{\bf r}}_i(t)}$.
The dynamic structure
factor ${\tilde S}({\bf k},\omega)$
of the charge density is defined as
\begin{equation}
   {\tilde S}({\bf k}, \omega) =
   {1 \over 2 \pi} \int^{+\infty}_{-\infty} {\rm d}t \,
    e^{-\imath \omega t} S({\bf k},t),
\label{S}
\end{equation}
\begin{eqnarray}
   S({\bf k},t) & = & {1 \over N} \left\langle
       \hat{\rho}_{\bf k}^\dagger(t) \hat{\rho}_{\bf k}(0) \right\rangle_T
       - N \delta_{{\bf k},0}
\nonumber \\
     & = & {1 \over N} \sum_{ij}
          \left\langle e^{\imath {\bf k} \cdot {\bf r}_i(t)}
          e^{-\imath{\bf k}\cdot {\bf r}_j(0)} \right\rangle_T
\nonumber \\
       & &  -(2 \pi)^3 n \delta({\bf k}),
\label{def-S}
\end{eqnarray}
where $N$ is the number of ions in the system, $n$ is the ion number density,
$\langle \ldots \rangle_T$ means canonical averaging at
temperature $T$, and the last term takes into account
contribution from the neutralizing background.

The above definition is equally valid for liquid and
solid states of the ion system. In the solid regime,
it is natural to set $\hat{{\bf r}}_i(t)= {\bf R}_i + \hat{\bf u}_i(t)$, 
where
${\bf R}_i$ is a lattice vector, and $\hat{\bf u}_i(t)$
is an operator of ion displacement from ${\bf R}_i$.
Accordingly,
\begin{eqnarray}
    S({\bf k},t) & = & {1 \over N} \sum_{ij}
    e^{\imath {\bf k} \cdot ({\bf R}_i - {\bf R}_j)} \,
    \left\langle e^{\imath {\bf k} \cdot \hat{\bf u}_i(t)} \,
    e^{ - \imath {\bf k} \cdot \hat{\bf u}_j(0)}
    \right\rangle_T
\nonumber \\
    & &  -(2 \pi)^3 n \delta({\bf k}).
\label{Sqt-det}
\end{eqnarray}

The main subject of the present paper is to discuss the
{\it harmonic
lattice} (HL) model which consists in
replacing the canonical averaging, $\langle \ldots \rangle_T$,
based on the exact Hamiltonian,
by the averaging based on the corresponding oscillatory Hamiltonian
which will be denoted as $\langle \ldots \rangle_{T0}$.
In order to perform the latter averaging we expand
$\hat{\bf u}_i(t)$ in terms of phonon normal coordinates:
\begin{eqnarray}
     \hat{\bf u}_i(t) & = &
     \sum_\nu \sqrt{ \hbar \over 2mN\omega_\nu}
     \,{\bf e}_\nu \times
\nonumber \\
     & & \left( e^{\imath{\bf q}\cdot {\bf R}_i - \imath\omega_\nu t}
        \, \hat{b}_\nu + e^{-\imath{\bf q}\cdot {\bf R}_i
        + \imath\omega_\nu t} \, \hat{b}_\nu^\dagger \right),
\label{u}
\end{eqnarray}
where $m$ is the ion mass,
$\nu \equiv ({\bf q},s)$,
$s=1,2,3$ enumerates phonon branches;
${\bf q}$, ${\bf e}_\nu$, $\omega_\nu$ are, respectively, 
phonon wavevector (in the first Brillouin zone),
polarization vector, and frequency; 
$\hat{b}_\nu$ and $\hat{b}^\dagger_\nu$
refer to phonon annihilation and creation operators.
The averaging over the oscillatory Hamiltonian,
$H_0 = \sum_\nu \frac{1}{2} \hbar \omega_\nu 
(\hat{b}_\nu \hat{b}_\nu^\dagger + \hat{b}_\nu^\dagger
\hat{b}_\nu)$, reads
\begin{equation}
    \langle \hat{F} \rangle_{T0} =
    \sum_\nu \sum_{n_\nu}^\infty f(n_\nu)F_{n_\nu n_\nu},
\label{T0}
\end{equation}
where $n_\nu$ is the number of phonons in a mode $\nu$,
$f(n_\nu)=e^{-n_\nu z_\nu} (1- e^{-z_\nu})$ is the phonon
density matrix in thermodynamic equilibrium,
$z_\nu= \hbar \omega_\nu/T$, $F_{n_\nu n_\nu}$
is a diagonal matrix element of the operator $\hat{F}$.
Inserting Eq.\ (\ref{u}) into (\ref{Sqt-det}) we can perform
the averaging (\ref{T0}) using the technique described, for instance,
in Kittel \cite{K63}.

The resulting structure factor $S({\bf k},t)$
takes into account absorption and emission of {\it any} number of phonons;
it can be decomposed into
the time-independent elastic (Bragg) part and
the inelastic part,
$S({\bf k},t) = S'({\bf k}) + S''({\bf k},t)$. The elastic part
is \cite{K63}:
\begin{equation}
    S'({\bf k}) = e^{-2W(k)} \, (2 \pi)^3 n \,
          {\sum_{\bf G}}' \delta({\bf k}-{\bf G}),
\label{S1}
\end{equation}
where ${\bf G}$ is a reciprocal lattice vector;
prime over the sum means
that the ${\bf G}=0$ term
is excluded (that is done due to the presence of uniform electron
background). 

In Eq.\ (\ref{S1}) we have introduced
the Debye-Waller factor,
$e^{-W(k)} = \left\langle \exp(\imath{\bf k} \cdot \hat{\bf u})
\right\rangle_{T0}$,
\begin{eqnarray}
    W(k) &=& {3 \hbar \over 2 m }  \left\langle
      {({\bf k}\cdot{\bf e}_\nu)^2 \over\omega_\nu}
         \left(\bar{n}_\nu+\frac12\right) \right\rangle_{\rm ph}
\nonumber \\
     & &
         =  { \hbar k^2 \over 2 m }
        \left\langle
      { 1 \over\omega_\nu}
         \left(\bar{n}_\nu+\frac12\right) \right\rangle_{\rm ph},
\label{DW}
\end{eqnarray}
where $\bar{n}_\nu =
\left( {\rm e}^{z_\nu}-1 \right)^{-1}$ is
the mean number of phonons in a mode $\nu$.
The brackets
\begin{equation}
\langle f_\nu \rangle_{\rm ph} = {1 \over 3N} \sum_\nu f_\nu \,
= { 1 \over 24 \pi^3 n} \sum_{s=1}^3 \int {\rm d}{\bf q} \, f_\nu
\label{ph}
\end{equation}
denote averaging over
the phonon spectrum, which can be performed numerically,
e.g., Ref.\ \cite{BY95}. The integral on the rhs is meant to be taken over the
first Brillouin zone.
The latter equality in Eq.\ (\ref{DW}) is exact
at least for cubic crystals discussed below.
For these crystals,
$W(k) = r_T^2 k^2/6$, where $r_T^2 = \langle \hat{\bf u}^2 \rangle_{T0}$
is the mean-squared ion displacement (e.g., \cite{K63,BY95}).

The inelastic part of $S({\bf k},t)$ (e.g., \cite{K63})
can be rewritten as
\begin{eqnarray}
     S''({\bf k},t) &=& \sum_{\bf R} \, e^{i{\bf k} \cdot {\bf R} -2 W(k)}
       \, \left[ e^{v_{\alpha \beta} ({\bf R},t)
       k_\alpha k_\beta} -1 \right],
\label{S2a} \\
   v_{\alpha \beta}({\bf R},t)
   &=& {3 \hbar \over 2m}
   \left\langle {e_{\nu \alpha} e_{\nu \beta} \over \omega_\nu}  \,
  {\cos{(\omega_\nu t + i z_\nu/2)} \over \sinh{(z_\nu/2)} }
   \, e^{i {\bf q}\cdot{\bf R}}
  \right\rangle_{\rm ph}.
\nonumber \\
\label{vab}
\end{eqnarray}
Eqs.\ (\ref{S1}) and (\ref{S2a}) result in the HL dynamical
structure factor
\begin{eqnarray}
     {\tilde S}({\bf k},\omega) &=&
      - (2 \pi)^3 n \, \delta(\omega) \delta({\bf k})
\nonumber \\
 &&
 + {1 \over 2 \pi}
     \int^{+\infty}_{-\infty} {\rm d}t \, e^{-i\omega t - \hbar \omega/2T} \,
 \nonumber \\
  &&\times
   \sum_{\bf R} e^{i {\bf k}\cdot{\bf R} -2 W(k) +
     v_{\alpha \beta}({\bf R}, \tau) k_\alpha k_\beta}~,
\end{eqnarray}
where $t$ is real and $\tau=t-i\hbar/(2T)$.

Along with the HL model
we will also use the simplified model introduced in Ref.\ \cite{BKPY98}.
It will be called HL1
and its results will be labelled by the subscript `1'.
It consists in replacing
$S''({\bf k},t)$ given by
Eq.\ (\ref{S2a}) by a simplified expression $S''_1({\bf k},t)$
equal to the first term of the sum, ${\bf R}=0$:
\begin{eqnarray}
       S_1({\bf k},t)& = & S'({\bf k})+S''_1({\bf k},t),
\nonumber \\
     S''_1({\bf k},t) &=&  e^{-2 W(k)}
       \, \left( e^{v(t) k^2} -1 \right),
\label{HL1eq}
\end{eqnarray}
where $v$ is defined by the equation
$v_{\alpha \beta}(0,t) = v(t) \, \delta_{\alpha \beta}$,
which is the exact tensor structure
for cubic crystals (see above).
%In this way the HL1 model neglects direct effect of correlations
%of different ions on the inelastic structure factor $S''_1({\bf k},t)$.
The accuracy of this approximation, as discussed in Ref.\ \cite{BKPY98},
is good for evaluating the quantities obtained by
integration over {\bf k}
(e.g., transport properties of degenerate electrons
in Coulomb crystals of ions).

%%%%%%%%%%%%%%%%%%%% SECTION 3 %%%%%%%%%%%%%%%%%%%%%%%%%%%%%%%%%%%
%%%%%%%%%%%%% STATIC CORRELATION FUNCTION %%%%%%%%%%%%%%%%%%%%%%%%
\section{Static case. HL versus MC}

In this section we compare our analytic models with
MC simulations of Coulomb crystals. For this purpose
we introduce the function
\beq
    g(r)= 1 + {1 \over n} \int
    {{\rm d}\Omega_{\bf r} \over 4 \pi}
    \int { {\rm d}{\bf k} \over (2 \pi)^3 } \,
    [S({\bf k},0)-1] \, e^{-i {\bf k} \cdot {\bf r}},
\label{gc}
\eeq
which may be called the static two particle radial distribution
function. This function is the result of an angular and a translation
average of the static two particle distribution function.
In this expression
d$\Omega_{\bf r}$ is the solid angle element in the direction of {\bf r}.
One can see that $4 \pi r^2 n g(r) {\rm d}r$ is the ensemble averaged
number of ions in a spherical shell of radius $r$ and width d$r$
centered at a given ion.
Thus $g(r)$ is just the quantity determined
from MC simulations \cite{SDS90}.

First let us use
the HL1 model. From
Eqs.\ (\ref{S1}) and (\ref{HL1eq})
we easily obtain $g_1(r) = g'(r) + g''_1(r)$, where
\begin{eqnarray}
       g'(r) &=& 1+{\sum_{\bf G}}' e^{-2W(G)}\,
       {\sin{Gr} \over Gr},
\nonumber \\
       g''_1(r) &=&
       - {3 \sqrt{3\pi} \over 8 \pi^2 n r^3_T}
       \, \exp \left( -{3 r^2 \over 4 r^2_T} \right).
\label{g1}
\end{eqnarray}

Calculation of $g''(r)$ in the HL model
is more cumbersome. After integration over
$k=|{\bf k}|$ and $\Omega_{\bf r}$ the result can be written as
\begin{eqnarray}
        g(r) &=& g_1(r)  +
        {\sum_{\bf R}}' \sum_{\sigma=\pm 1} \left[
        {\sqrt{\pi} \over (2\pi)^3 r n}
 \right.
 \nonumber \\
        && \times  \left.
        \int {{\rm d}\Omega_{\bf k} \over x^2}
         \gamma \, e^{-\gamma^2}
         + {\sqrt{3 \pi} \sigma \over 8 \pi^2 n r R r_T} \,
          e^{-\eta} \right],
\label{gr}
\end{eqnarray}
where $\gamma=(r+\sigma R \mu)/x$, $\eta=3(r+\sigma R)^2/(4 r_T^2)$,
$\mu= \cos \vartheta$, $\vartheta$ is an angle between
{\bf k} and {\bf R},
$x^2 = 4 [r^2_T/3 -  (k_\alpha k_\beta v_{\alpha \beta}({\bf R},0) / k^2)]$,
and d$\Omega_{\bf k}$ is the solid angle element in the direction of {\bf k}.
Therefore, we need to evaluate a rapidly
converging lattice sum (\ref{gr}) of 2D integrals in which $x$
is known once the matrix
elements $v_{\alpha \beta}({\bf R},0)$
are calculated from Eq.\ (\ref{vab}). 
We have performed
the integration over the first Brillouin zone required in
Eq.\ (\ref{vab}) using the 3D Gauss integration scheme
described in Ref.\ \cite{AGpre81}.

The function $g(r)$ depends on the lattice type and on two parameters:
the classical ion coupling parameter $\Gamma = Z^2 e^2/(aT)$
and the quantum parameter $\theta = \hbar \omega_p/T$
that measures the importance of zero-point lattice vibrations.
In this case $Ze$ is the ion charge,
$a=(4 \pi n/3)^{-1/3}$ is the ion sphere radius, and
$\omega_p=Ze \, \sqrt{4 \pi n / m}$ the ion plasma frequency.

\begin{figure}[t]
\begin{center}
\leavevmode
\epsfysize=8.5cm
\epsfbox{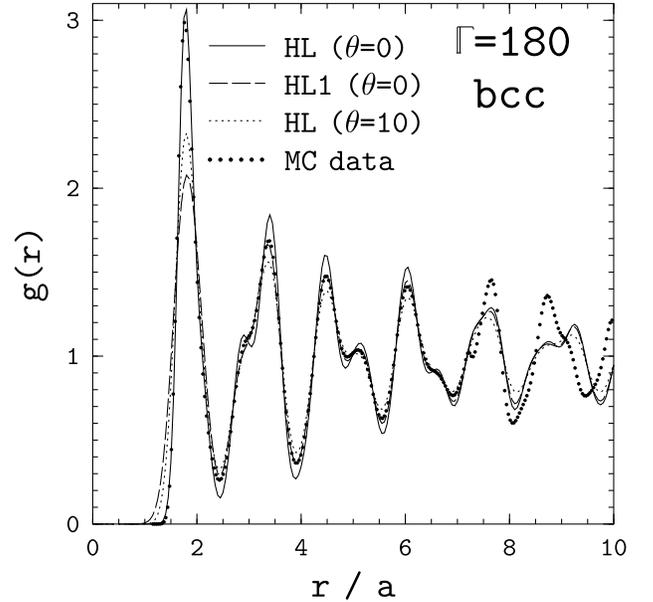}
\end{center}
\caption[ ]{$g(r)$ for a bcc Coulomb crystal at $\Gamma=180$.}
\label{fig1}
\end{figure}

\begin{figure}[t]
\begin{center}
\leavevmode
\epsfysize=8.5cm
\epsfbox{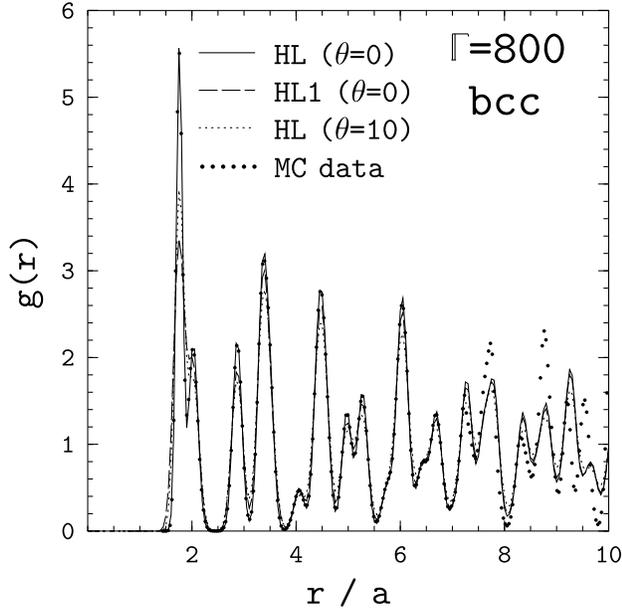}
\end{center}
\caption[ ]{Same as in Fig.\ \ref{fig1} but at $\Gamma=800$.}
\label{fig2}
\end{figure}

First consider a classical Coulomb crystal, $\theta \to 0$,
for which
$\bar{n}_\nu \approx T/(\hbar \omega_\nu)$.
The functions $g(r)$ calculated
using the HL and HL1 models for body-centered cubic (bcc) crystals
at $\Gamma$ = 180 and 800 are
presented in Figs.\ \ref{fig1} and \ref{fig2}.
The pronounced peak structure corresponds to the bcc lattice vectors.
These results are compared with extensive MC simulations.
The MC method is described, e.g., in Ref.\ \cite{SDS90}.
The simulations have
been done with 686 particles over nearly $10^8$
MC configurations.

%                                                       FIGURE fig1

One can observe a very good agreement of HL and MC results
for both values of $\Gamma$ at $1.5 \lesssim r/a \lesssim 7$.
The MC results for $g(r)$ are limited to half the size of the basic cell
containing the $N$ charges due to the bias from particles in the image cells
adjacent to the basic cell. For $N = 686$ the basic cell length is 14.2 $a$.
Hence the MC $g(r)$ results for this simulation are valid only out to
$r \approx 7\,a$ while $g(r)$, given by the HL model, 
remains accurate as $r \to \infty$.
At small particle separations, $r \lesssim 1.5\,a$,
where $g(r)$ becomes small, the HL $g(r)$ deviates from the MC $g(r)$.
It is clear that the HL model cannot be reliable at
these $r$, where strong Coulomb repulsion of two particles
dominates, and the MC data (available down to $r \gtrsim 1.1\,a$)
are more accurate.
The HL1 model is quite satisfactory at
$r \gtrsim 2.5\,a$, beyond the closest lattice peak.
The HL model improves significantly HL1 at lower $r$.
It is interesting that
for $\Gamma=180$ the HL1 model agrees slightly better with MC for the
range $2.5 \lesssim r/a \lesssim 6$ than the HL model does.
With increasing $\Gamma$, however, the HL model comes into better
agreement with MC at these $r$, although the difference between the  
HL and HL1 models
becomes very small. This good agreement of the HL models
with the MC simulations after the first peak of $g(r)$ 
indicates that we have
a very good description of Coulomb crystals for which the HL model may be
used in place of MC simulations.

The HL model enables one to analyse quantum effects.
Figs.\ \ref{fig1} and \ref{fig2} exhibit also
$g(r)$ in the quantum
regime at $\theta=10$.
Zero-point lattice vibrations
tend to reduce lattice peaks. The simplicity of the implementation of the
HL model in the quantum regime is remarkable given the complexity of
direct numerical studies of the quantum
effects by MC, PIMC or MD simulations
(see, e.g., Ref.\ \cite{O97}).

%                                                       FIGURE fig2
%

%%%%%%%%%%%%%%%%%%%%%%%%  Section 4      %%%%%%%%%%%%%%%%%%%%%%%%%%%%%%%
%%%%%%%%%%%%%%%%%%%%%  POTENTIAL ENERGY  %%%%%%%%%%%%%%%%%%%%%%%%%%%%%%%
\section{Coulomb energy}

To get a deeper insight into the HL and HL1 models
let us use them to
calculate the electrostatic energy $U$
of the crystal. Writing this energy as the sum
of Coulomb energies of different pairs of ions complemented by
the interaction energy of ions with the electron background and
the Coulomb energy of the background itself, we arrive at the
standard expression
\begin{equation}
  {U \over N} = 2 \pi n  \int^\infty_0 r^2 \, {\rm d}r \,
  {Z^2 e^2 \over r} \, [g(r)-1],
\label{Def}
\end{equation}
where $g(r)$ is given by Eq.\ (\ref{gc}). 
Therefore, we can
use the function $g(r)$ calculated in Sect.\ 3 to analyse $U$.

For the HL1 model from Eqs.\ (\ref{g1}) we get
\begin{eqnarray}
&&
       {U_1 \over N T} =
      {\sum_{\bf G}}' e^{-2W(G)} \, {2 \pi n Z^2 e^2 \over T G^2}
      - \sqrt{3 \over 4 \pi}{Z^2 e^2 \over T r_T} =
\nonumber \\
&&
       \Gamma \left[ \zeta + {r_T^2 \over 2 a^2} -
      {\sum_{\bf R}}'  {a \over 2R} \,
      {\rm erfc}\left({\sqrt{3} R \over 2 r_T}\right) \right],
\label{U1}
\end{eqnarray}
where $\zeta$ is the electrostatic Madelung constant
[$=-0.895929$ for bcc, and
$-0.895873$ for face-centered cubic (fcc)
lattice], and ${\rm erfc}(x)$ is the complementary error function.
The second line of this equation is obtained using
the formula for the Madelung constant
derived with the Ewald method (see, e.g., Ref.\ \cite{BH54}) 
\begin{eqnarray}
   \zeta &=& 
            {\sum_{\bf R}}' {a \over 2R} \, 
               {\rm erfc}\left({AR\over a} \right) +
             {3 \over 2}
             {\sum_{\bf G}}'
              {e^{- G^2 a^2 / (4 A^2)} \over G^2 a^2} 
\nonumber\\
        &&   -  {3 \over 8 A^2} -
              {A \over \sqrt{\pi}}~,
\label{fuchs}
\end{eqnarray}
where $A$ is an arbitrary number. In the particular case of
Eq.\ (\ref{U1}) $A = \sqrt{3} a / (2 r_T)$. 

For the HL model, using Eq.\ (\ref{gr}), we have
\begin{eqnarray}
&&
  {U \over N T } =
  \Gamma \left\{ \zeta + {r_T^2 \over 2 a^2}
  \right.
\nonumber \\
&&
    -  \left. {\sum_{\bf R}}' \left[ {a \over 2 R} -
    \int {{\rm d}\Omega_{\bf k} \over 4 \pi^2} {\sqrt{\pi} a \over x}
    \exp \left(-{R^2 \mu^2 \over x^2} \right) \right] \right\}.
\label{U}
\end{eqnarray}

First, consider the classical crystal at zero temperature, $T \to 0$.
Then $r_T \to 0$, $x \to 0$, and
we reproduce the Madelung energy,
$U/N \to U_1/N \to \zeta Z^2 e^2/a$.
In the limit of small $T$ both $U_1/N$ and $U/N$ contain
the main term that can be expanded
in powers of $T$ plus an exponentially small term (non-analytic at $T=0$).
For the classical
crystal at any $T$ we have $r_T^2/a^2 = u_{-2}/\Gamma$, where
$u_s = \langle (\omega_\nu / \omega_p)^s \rangle_{\rm ph}$
denotes a phonon spectrum moment
($u_{-2}$=12.973 for bcc and 12.143 for fcc).

The sum over ${\bf R} \neq 0$ in the last expression for $U_1$
in Eq.\ (\ref{U1}) is exponentially small.
Thus the analytic part of $U_1$ in the HL1 model is given
only by two terms,
$U_1/(NT) = \zeta \Gamma + u_{-2}/2$. We see that the HL1 model fails to
reproduce correctly the harmonic part of the potential energy:
$u_{-2}/2$ appears instead of conventional $3/2$. 

On the contrary, the expansion of $U/(NT)$ in the HL model, Eq.\ (\ref{U}),
contains all powers of $T$.
To analyse this expansion, let us take any term
of the sum over ${\bf R}$, and introduce a local
coordinate frame with $z$-axis along ${\bf R}$.
Then
\begin{equation}
    \int \,{\rm d}\Omega_{\bf k} \,\ldots =
    \int_{-1}^{+1} \, {\rm d}\mu \,
    \int_0^{2 \pi} \, {\rm d}\phi \, \ldots,
\label{integral}
\end{equation}
where $\phi$ is an azimuthal angle of ${\bf k}$ in the adopted
frame. Since $x \to 0$
as $T \to 0$ in the denominator of the exponent
under the integral in Eq.\ (\ref{U}), only a narrow
interval of $\mu$ in the vicinity of $\mu=0$ contributes,
and we can extend the integration over $\mu$ to the interval
from $-\infty$ to $+\infty$. Furthermore, using the
definition of $x$, Eq.\ (\ref{gr}), we can
rewrite $x$ as
\begin{eqnarray}
  x^2 & = & x_0^2 \,(1+\epsilon), \quad
  \epsilon = {x_\mu^2 \over x_0^2},
\label{x2} \\
  x_0^2 & = & {4 \over 3} r_T^2 -
    4 \left( v_{xx} \cos^2 \phi + v_{yy} \sin^2 \phi
       +  v_{xy} \sin 2 \phi \right),
\nonumber \\
  x_\mu^2 & = & 4 \mu^2 \left(
      v_{xx} \cos^2 \phi + v_{yy} \sin^2 \phi
      + v_{xy} \sin 2 \phi - v_{zz} \right)
\nonumber \\
  & & - 8 \mu \, \sqrt{1 - \mu^2}
       \left( v_{xz} \cos \phi + v_{yz} \sin \phi \right),
\nonumber
\end{eqnarray}
where $v_{\alpha \beta}= v_{\alpha \beta}({\bf R},0)$.
Accordingly, we can treat $\epsilon$ as small
parameter and expand any integrand in Eq.\ (\ref{U})
in powers of $\epsilon$ and further in powers of $\mu$. 
This generates the
expansion in powers of $T$. 

We have been able to evaluate
three first terms of this expansion.
In particular, the term linear in $T$
contains the expression
\begin{eqnarray}
&&
     {3 T \over 2} \left\langle {\omega_p^2 \over \omega^2_\nu}
     {1 \over 4 \pi n} {\sum_{\bf R}}'
     {R^2 - 3 ({\bf R}\cdot{\bf e}_\nu)^2 \over R^5} \,
     e^{i {\bf q}\cdot{\bf R}} \right\rangle_{\rm ph}
\nonumber \\
&&
      =
     {3 T \over 2} \left\langle {\omega_p^2 \over \omega^2_\nu}
     \left[
     {\cal D}_{\alpha \beta}({\bf q}) e_{\nu\alpha} e_{\nu\beta} -
     {1\over 3} \right] \right\rangle_{\rm ph},
\label{DyM}
\end{eqnarray}
where ${\cal D}_{\alpha \beta}$ is the dynamical matrix.
Combining this expression with $r_T^2/(2a^2)$
and taking into account that
${\cal D}_{\alpha \beta} e_{\nu \alpha} e_{\nu \beta} =
\omega^2_\nu/\omega_p^2$
(according to the basic equation for the phonon spectrum)
we see that the HL expansion of
the analytic part of $U$ in powers of $T$
is $U/(NT)= \zeta \Gamma + 3/2 + \delta U_T/(NT)$; it
reproduces not only the Madelung term, but also
the correct oscillatory term $3/2$, and contains
a higher-order contribution $\delta U_T/(NT) =
A_1^{\rm HL} /\Gamma + A_2^{\rm HL} /\Gamma^2 + \ldots$
that can be called ``anharmonic'' contribution in the HL model.
After some transformations the coefficients 
$A_1^{\rm HL}$ and $A_2^{\rm HL}$ are reduced
to the sums over {\bf R} containing, respectively, bilinear and
triple products of $v_{\alpha \beta}$
(with integration over $\mu$ and $\phi$ done
analytically). Numerically the sums yield 
$A_1^{\rm HL}=10.64$ and $A_2^{\rm HL}=-62.4$.

The anharmonic terms occur since $U$, as given by Eq.\ (\ref{Def}),
includes exact Coulomb energy
(without expanding the Coulomb potential in powers of
ion displacements {\bf u}). However, we use $g(r)$ in the HL
approximation and thus neglect
the anharmonic contribution in ion-ion correlations.
Therefore, the HL model does not include all anharmonic effects.

Let us compare the HL calculation of $\delta U_T$ with the exact
calculation of the first anharmonic term in the Coulomb energy of
classical Coulomb crystals by Dubin \cite{D90}.
The author studied the expansion
$\delta U^{\rm exact}_T/(NT) =
A_1^{\rm exact} /\Gamma + A_2^{\rm exact} /\Gamma^2 + \ldots$
and expressed the first term as
\begin{equation}
       A_1^{\rm exact} = \Gamma \left[
       {\langle U^2_3 \rangle \over 72 N T^2} -
       {\langle U_4 \rangle \over 24 N T }  \right]~,
\label{A1ex}
\end{equation}
where $U_n/n!$ is the $n$th term
of the Taylor expansion of the Coulomb energy over ion displacements,
while angular brackets denote
averaging with the harmonic Hamiltonian $H_0$.
According to Dubin $A_1^{\rm exact}$=10.84 and 12.34
for bcc and fcc crystals, respectively.
(The same quantity was computed earlier by Nagara et al.\ \cite{NNN87} 
who reported $A_1^{\rm exact}$=10.9 for bcc.)

It turns out that our $\delta U_T$ sums up
a part of the infinite series of anharmonic corrections to
the energy, denoted by Dubin as
$\sum_{n=3}^\infty \langle U_n\rangle/(n!)$, so that
$A_1^{\rm HL}= \Gamma \langle U_4 \rangle / (24 N T)$,
$A^{\rm HL}_2 = \Gamma^2 \langle U_6 \rangle / (6! N T)$, etc.
(The fact that this summation can be performed in a closed analytic form
was known from works on the so called
self-consistent phonon approximation, e.g., \cite{AG81} and references
therein.) Our numerical value
for the bcc lattice $A_1^{\rm HL}=10.64$
is very close to the
value of $\Gamma \langle U_4 \rangle / (24 N T)$
reported by Dubin as $\approx 10.69$
(his Table 3) which confirms accuracy of both calculations.
The fact that $A_1^{\rm HL}=10.64$ is close
to $A_1^{\rm exact}=10.84$ for bcc
is accidental (Dubin found $\Gamma
\langle U^2_3 \rangle / (72 N T^2) \approx 21.53$ for bcc). 
For instance,
from the results of Ref.\ \cite{D90} for fcc one infers,
$A_1^{\rm HL} \approx 5.63$ which differs strongly
from the exact anharmonic coefficient $A_1^{\rm exact}=12.34$.

Now let us set $T=0$ and analyse the quantum effects.
We can expand Eqs.\ (\ref{U1}) and (\ref{U}) in powers of $r_T/a$.
For $T=0$ the quantity $r_T$ tends to the rms amplitude
of zero-point vibrations,
$r_T=\sqrt{3 \hbar u_{-1}/(2 m \omega_p)}$, where
$u_{-1}$ is another phonon spectrum moment (=2.7986 and
2.7198 for bcc and fcc, respectively). The expansion of
$U_1/N$ gives $\zeta Z^2e^2/a + u_{-1} \hbar \omega_p /4$ plus
small non-analytic terms.
In the same manner as in Eq.\ (\ref{DyM})
we find that $U/N =\zeta Z^2e^2/a + 3 u_1 \hbar \omega_p/4 + \delta U_0/N$.
The second term gives half of the total
(kinetic + potential)
zero-point harmonic energy
of a crystal, as required by the virial theorem for
harmonic oscillator
($u_1$ =0.51139 and
0.51319 for bcc and fcc, respectively),
while the third term, $\delta U_0$, represents zero-point anharmonic
energy in the HL approximation.

To make the above algebra less abstract
let us estimate the accuracy of the HL
model and the relative importance of the anharmonicity
and quantum effects.
In the classical case,
taking $\Gamma=170$
(close to the melting value $\Gamma_{\rm m}=172$ for bcc),
we estimate the anharmonic contribution to
the total electrostatic energy as $|\delta U_T / U | \approx
A_1^{\rm exact}/(|\zeta|\Gamma^2) \approx
4.2 \times 10^{-4}$ and $4.8 \times 10^{-4}$ for bcc and
fcc, respectively.

The relative error into $U$
introduced by using the HL model is $A_2^{\rm exact}/(|\zeta|\Gamma^3)
\approx 5.7 \times 10^{-5}$ for bcc (if we adopt an estimate
of $A_2^{\rm exact} \approx 247$
from the MD data on the full electrostatic energy
presented in Table 5
of Ref.\ \cite{FH93})
and $[A_1^{\rm exact}-A_1^{\rm HL}]/(|\zeta|\Gamma^2) \approx
2.6 \times 10^{-4}$ for fcc. We see that Coulomb crystals
can be regarded as highly harmonic, and the accuracy of
the HL model is sufficient for many practical applications.
Obviously, the accuracy becomes even better with decreasing $T$.
The quantum effects can be more important (than the anharmonicity)
in real situations. Let us take
$^{12}$C matter at density
$\rho=10^6$ g cm$^{-3}$ typical for the white dwarf cores
or neutron star crusts.
The quantum contribution into energy is measured by
the ratio $3 u_1 \hbar \omega_p/(4 |\zeta| Z^2 e^2/a)$ which is equal to
$4.7 \times 10^{-3}$ at given $\rho$ (and
grows with density as $\rho^{1/6}$).

For completeness we mention that
the compressibility of the electron background
also contributes to the electrostatic energy.
The relative contribution
in the degenerate electron case
for $^{12}$C at $\rho=10^6$ g cm$^{-3}$ is $\sim 10^{-2}$
(e.g., Ref.\ \cite{CP98}).
Another point is that the HL model takes into account
zero-point lattice vibrations but neglects ion exchange
which becomes important at very high densities (e.g., Ref.\ \cite{C93}).

%%%%%%%%%%%%%%%%%%%%%%%%%%% Section 5 %%%%%%%%%%%%%%%%%%%%%%%%%%%%%%%%%%
%%%%%%%%%%%%%%%%%%%%%%% STATIC STRUCTURE FACTOR %%%%%%%%%%%%%%%%%%%%%%%%
\section{Structure factors}

\begin{figure}[t]
\begin{center}
\leavevmode
\epsfysize=8.5cm
\epsfbox{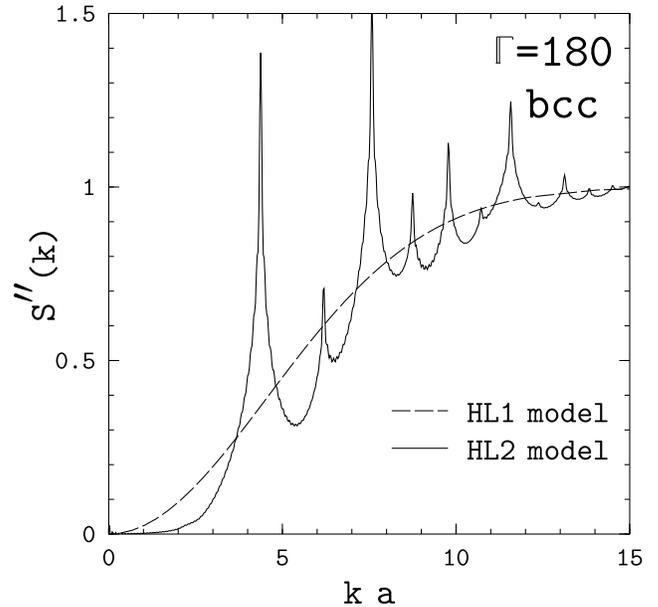}
\end{center}
\caption[ ]{Inelastic part of the structure factor at $\Gamma=180$
for classical bcc crystal.}
\label{fig3}
\end{figure}

Finally, it is tempting to use the HL model for analyzing
the ion structure factors themselves. Consider the angle-averaged static
structure factor $S(k)= \int {\rm d}\Omega_{\bf k} S({\bf k},t=0)/(4 \pi)$.
For the Bragg part, from Eq.\ (\ref{S1}) we obtain the expression
\begin{equation}
    S'(k) = e^{-2W(k)} \, 2 \pi^2 n \,
          {\sum_{\bf G}}' \delta(k-G)/G^2,
\label{S11}
\end{equation}
containing delta-function singularities at $k=G$, lengths of reciprocal
lattice vectors ${\bf G}$.
Direct HL calculation of $S''(k)$ from Eq.\ (\ref{S2a})
is complicated by the slow convergence of the sum
and complex dependence of $v_{\alpha\beta}$ on $R$.
However, the main features of $S''(k)$ can be
understood from two approximations.
First, in the HL1 model we have
$v_{\alpha \beta}(0,0) k_\alpha k_\beta = 2W(k)$, and
$S''_1(k)=1-e^{-2W(k)}$ as shown by
the dashed line in Fig.\ \ref{fig3}.

The second, more realistic approximation will be called HL2
(and labelled by the subscript `2').
It consists in
adopting a simplified tensor decomposition of
$v_{\alpha \beta}({\bf R},0)$ of the form
$v_{\alpha \beta}({\bf R},0) = F(R) \, \delta_{\alpha \beta} +
R_\alpha R_\beta J(R)/R^2$.
If so, we can immediately take the following integrals
$\int {\rm d}\Omega_{\bf R} \, v_{\alpha \alpha}({\bf R},0)/(4\pi)=
3F(R)+J(R)$
and $\int {\rm d}\Omega_{\bf R} \, v_{\alpha \beta}({\bf R},0)
R_\alpha R_\beta /(4 \pi R^2)=F(R)+J(R)$ (assuming summation
over repeating tensor indices $\alpha$ and $\beta$).
On the other hand, we can calculate the same integrals
taking $v_{\alpha \beta}({\bf R},0)$
from Eq.\ (\ref{vab}) at $t=0$.
In this way we come to
two linear equations for $F(R)$ and $J(R)$. Solving them, we
obtain
\begin{eqnarray}
    F(R) & = & { 3 \hbar \over 2m} \,
          \left\langle { 1 \over \omega_\nu}
          \left( \bar{n}_\nu + {1 \over 2} \right)
          \,  \biggl\{ j_0(y) - {j_1(y) \over y} \biggr. \right.
\nonumber \\
        & & \left. \left. -
     { ({\bf q} \cdot {\bf e}_\nu)^2 \over q^2} \left[j_0(y)
     - {3j_1(y) \over y} \right]
     \right\} \right\rangle_{\rm ph},
\nonumber \\
    J(R) & = & { 3 \hbar \over 2m} \,
          \left\langle {1 \over \omega_\nu}
          \left( \bar{n}_\nu + {1 \over 2} \right)
          \left[j_0(y) - {3j_1(y) \over y} \right] \right.
\nonumber \\
     & &    \times  \left.  \left[
     { 3 ({\bf q} \cdot {\bf e}_\nu)^2 \over q^2} -1
     \right] \right\rangle_{\rm ph},
\label{FJ}
\end{eqnarray}
where $y=qR$, and
$j_0(y)$ and $j_1(y)$ are the spherical Bessel functions.
Note that $F(0)k^2=2W(k)$, $J(0)=0$. 

In the limit of large $R$ the functions $j_0(qR)$ and
$j_1(qR)$ in Eqs.\ (\ref{FJ}) strongly oscillate which means
that the main contribution into the phonon averaging
(integration over ${\bf q}$) comes from a small vicinity
near the center of the Brillouin zone. Among three branches of
phonon vibrations in simple Coulomb crystals, two ($s$=1, 2) behave
as transverse acoustic modes, while the third ($s$=3) behaves as
a longitudinal optical mode ($\omega \approx \omega_p$)
near the center of the Brillouin zone. Owing to the
presence of $\omega_\nu^{-1}$ in the denominator of Eqs.\ (\ref{FJ}),
the main contribution at large $R$ comes evidently from the acoustic modes.
Thus we can neglect optical phonons and
set $\omega=c_s q$ for acoustic modes, where $c_s$ is the mean
ion sound velocity. In the high-temperature classical limit,
$(\bar{n}_\nu + \frac{1}{2}) \to T/(\hbar c_s q)$.
Then from Eqs.\ (\ref{FJ}) at $R \to \infty$ we approximately
obtain
\begin{eqnarray}
 F(R) & \approx & {T \over 4 \pi^2 n  m R }
       \int_0^\infty {\rm d}y \,\left[ j_0(y)-{j_1(y) \over y} \right]
       \sum^2_{s=1} {1 \over c_s^2}
\nonumber \\
    &= & { T \over 16 \pi n m R} \sum^2_{s=1} {1 \over c_s^2}~,
\nonumber \\
 J(R) & \approx & - {T \over 4 \pi^2 n m R }
       \int_0^\infty {\rm d}y \,\left[ j_0(y)-{ 3j_1(y) \over y} \right]
       \sum^2_{s=1} {1 \over c_s^2}
\nonumber \\
     &  = & { T \over 16 \pi n m R} \sum^2_{s=1} {1 \over c_s^2}.
\label{FJasy}
\end{eqnarray}
Our analysis shows that an appropriate
value of $c_1^{-2}+ c_2^{-2}$ for bcc lattice
would be $67.85/(a \omega_p)^2$.
From Eq.\ (\ref{FJasy}) we see that $F(R)$ and $J(R)$ decrease
as $R^{-1}$ with increasing $R$. In the quantum limit $\theta \gg 1$
we have
$(\bar{n}_\nu + \frac{1}{2}) \to \frac{1}{2}$; applying the same
arguments we deduce that $F,J \propto R^{-2}$ as $R \to \infty$.

Using Eq.\ (\ref{S2a})
we have
\begin{eqnarray}
   S''_2(k)& = & \int {{\rm d}\Omega_{\bf k} \over 4 \pi} \,
       \sum_{\bf R} e^{\imath {\bf k}\cdot {\bf R}-2W(k)}
\nonumber \\
    &  & \times \left[ e^{k^2F(R)+({\bf k} \cdot {\bf R}/R)^2 J(R)}-1
\right]
      =  1 - e^{-2W(k)}
\nonumber \\
     & & + ~{1 \over 2}~ {\sum_{\bf R}}' \int_{-1}^{+1}
     {\rm d}\mu \, e^{-2W(k)+\imath kR \mu}
\nonumber \\
     & & \times   \left[ e^{k^2 F(R)+k^2 J(R) \mu^2}-1 \right] .
\label{S22}
\end{eqnarray}
A number of the first terms of the sum, say for $|{\bf R}| < R_0$, 
where $R_0/a$ 
is sufficiently large, can be calculated exactly.
To analyse the convergence of the sum over ${\bf R}$ at large $R$
let us expand 
the exponential in the square brackets on the rhs. All the terms
of the expansion which  behave as $R^{-n}$ with 
$n \ge 2$ lead to nicely convergent contributions to $S''_2(k)$.
The only problem is posed by the linear expansion term
in the {\it classical} case. The tail of the sum, 
$\sum_{|{\bf R}|>R_0}$, 
for this term can be regularized and calculated by the Ewald method
(e.g., Ref.\ \cite{BH54})
with the following result
\begin{eqnarray}
    && \int {{\rm d} \Omega_{\bf k} \over 4 \pi}  
    \sum_{|{\bf R}|>R_0} e^{\imath {\bf k} \cdot {\bf R} - 2 W(k)}
    \left[ e^{k^2F+({\bf k} \cdot {\bf R}/R)^2 J}-1 \right]
\nonumber \\
   &\approx&  { 2T k^2 e^{-2W(k)} \over 16 \pi n m} \sum^2_{s=1} {1 \over c_s^2} 
        \left[\sum_{|{\bf R}|>R_0} {\sin{kR} \over kR^2} \, 
         \, {\rm erfc} \left({AR\over a}\right) 
         \right.
\nonumber \\
   &&+  \left.
         {4 \pi n \over k^2} \, e^{-k^2 a^2/(4 A^2)} +
          {\sum}'_{|{\bf R}|<R_0} {\sin{kR} \over kR^2} \, 
          \, {\rm erf} \left({AR\over a} \right)
          \right.
\nonumber \\
       && + \left. {\sum_{\bf G}}' \sum_{\tau = \pm 1} {\pi n \tau \over kG}
         \, {\rm Ei}\left(-{[k+ \tau G]^2 a^2 \over 4 A^2} \right)
         + {2A \over a \sqrt{\pi}}
       \right],    
\label{tail-sofist}
\end{eqnarray}
where ${\rm Ei}(-x)$ is the exponential integral, and $A$ is a 
number to be chosen in such a way 
the convergence of both infinite sums (over direct and reciprocal lattice
vectors) be equally rapid. 
Letting $A \to \infty$ we obtain a much more
transparent, although slower convergent formula
\begin{eqnarray}
    \Bigg[ \ldots \Bigg] &=& {4 \pi n \over k^2} +  2 \pi n {\sum_{\bf G}}' 
       \left[{1 \over kG}
    \ln{\left| {k+G \over k-G} \right|} - {2 \over G^2} \right]
\nonumber \\
    && - {\sum}'_{|{\bf R}|<R_0}  {\sin{kR} \over k R^2} + {2 \zeta \over a}~.
\label{tail-ex}
\end{eqnarray}

This expression explicitly reveals logarithmic singularities at $k=G$.
They come from inelastic processes of
one-phonon emission or
absorption in the cases in which given wave vector ${\bf k}$ is close
to a reciprocal lattice vector ${\bf G}$.
To prove this statement let us perform Taylor expansions
of both exponentials in angular brackets in
Eq.\ (\ref{Sqt-det}).
The one-phonon processes correspond to
those expansion terms 
which contain products of one creation
and one annihilation operator. Thus, 
in the one-phonon approximation
$S''({\bf k},t=0)$ reads
\begin{eqnarray}
& & S''_{\rm 1ph}({\bf k},t=0)  =  {e^{-2 W(k)} \over N} \sum_{ij} 
       e^{\imath {\bf k} \cdot ({\bf R}_i -{\bf R}_j)} 
\nonumber \\       
    & & \times   \left\langle 
            (i{\bf k} \cdot \hat{\bf u}_i)
            (-i{\bf k} \cdot \hat{\bf u}_j) \right\rangle_{T0}
\nonumber \\
& &= {e^{-2 W(k)} \over N}\sum_{ij} 
  \sum_\nu {\hbar ({\bf k} \cdot {\bf e}_\nu)^2 \over 2 m N \omega_\nu} 
 \, e^{\imath ({\bf k} - {\bf q}) \cdot ({\bf R}_i - {\bf R}_j)} 
            (2 \bar{n}_\nu+1) 
\nonumber \\
 &&= e^{-2 W(k)} \sum_s  {\hbar ({\bf k} \cdot {\bf e}_{{\bf q}s})^2 \over  m
     \omega_{{\bf q}s}} \left(\bar{n}_{{\bf q}s}+
       \frac{1}{2} \right), 
\label{1ph}
\end{eqnarray}
where the last summation is over phonon polarizations,
${\bf q}={\bf k}-{\bf G}$ is the phonon wave vector
which is the given wave vector ${\bf k}$
reduced
into the first Brillouin zone by subtracting an appropriate
reciprocal lattice vector ${\bf G}$.
In addition, in
Eq.\ (\ref{1ph}) we have introduced an overall factor
$e^{-2W(k)}$ which comes from renormalization of the one-phonon
probability associated with
emission and absorption of any number of virtual
phonons (e.g., Ref.\ \cite{K63}).
Now let us assume that $|k-G|a \ll 1$ and average
Eq.\ (\ref{1ph}) over orientations of ${\bf k}$
[integrate over ${\rm d}\Omega_{\bf k}/(4 \pi)$].
One can easily see that the important contribution
into the integral comes from a narrow cone
$\Omega_0$ aligned along ${\bf G}$.
Let $\theta_0 \ll 1$ be the cone angle
chosen is such a way that $G \theta_0 a \ll 1$, but 
$G\theta_0 \gg |G-k|$. 
Integrating within this
cone, we can again adopt approximation of
acoustic and longitudinal phonons and neglect
the contribution of the latters. For simplicity,
we also assume that the sound velocities of
both acoustic branches are the same:
$\omega_\nu = c_s |{\bf k} - {\bf G}|$.
Then, in the classical limit
we come to the integral of the type
\begin{equation}
  \int_{\Omega_0} {{\rm d} \Omega_{\bf k} \over 4 \pi}
      \sum_{s=1}^2 { ({\bf k} \cdot {\bf e}_{{\bf q}s})^2 \over
      \omega_{{\bf q}s}^2}  \approx
      {1 \over 4 c_s^2} \left\{ \ln \left[ {k G \theta_0^2
         \over (k-G)^2} \right] - 1 \right\},
\label{singul}
\end{equation}
which contains exactly the same logarithmic divergency
we got in Eq.\ (\ref{tail-ex}). Note that in the quantum
limit we would have similar integral but with
$\omega$ instead of $\omega^2$ in the denominator
of the integrand. The integration would yield
the expression proportional to $|k-G|$, i.e.,
the logarithmic singularity would be replaced by
a weaker kink-like feature. Therefore,
the $k=G$ features of the inelastic structure
factor $S''(k)$ in the quantum limit are expected to be
less pronounced than in the classical limit but
could be, nevertheless, quite visible. Actually, at any finite temperature,
even deep in the quantum regime $T \ll \hbar \omega_p$ 
there are still phonons excited thermally 
near the very center of the Brillouin zone, where the energy of acoustic
phonons is smaller than temperature. Due to these phonons the logarithmic 
singularity 
always exists on top of the kink-like feature at $T \ne 0$.

After this simplified consideration let us return
to qualitative analysis.
We have calculated $S''_2(k)$
in the classical limit using the HL2 approximation
as prescribed above and verified
that the result is indeed independent of $R_0$ 
(in the range from $\sim 30a$ to $100a$) and $A$.
The resulting $S''_2(k)$
is plotted in Fig.\ \ref{fig3} by the solid line.

%                                                       FIGURE fig3

Thus, in a crystal, the inelastic part of the structure factor,
$S''(k)$, appears to be singular in addition to
the Bragg (elastic) part $S'(k)$. 
The singularities of $S''(k)$ are weaker
than the Bragg diffraction
delta functions in $S'(k)$; the positions of singularities
of both types coincide.
The pronounced shapes of the $S''(k)$ peaks may, in principle,
enable one to observe them experimentally. 
The structure factor $S(k)$ in the Coulomb liquid 
(see, e.g., Ref.\ \cite{YCDW91} and references therein)
also contains significant but finite and regular humps associated
with short-range order. This structure has been studied in detail
by MC and other numerical methods.  
In contrast, the studies of singular structure factors in a crystal
by MC or MD methods would be very complicated. Luckily,
they can be explored by the HL model.
 
Finally, it is instructive to compare the behavior of
$S''(k)$ at small $k$
in the HL1 and HL2 models. It is easy to see that
the main contribution to inelastic
scattering at these $k$ comes from one-phonon {\it normal}
processes [with {\bf q}={\bf k} in Eq.\ (\ref{1ph})].
At these $k$ the HL2 $S''_2(k)$ coincides with  
the one-phonon
$S''_{\rm 1ph}(k)$  and
with the static structure factor of Coulomb liquid 
(at the same $\Gamma$) and
reproduces correct
hydrodynamic limit \cite{VH75}, $S(k) \propto k^2$.
The HL1 model, on the contrary, overestimates the importance
of the normal processes.

Let us mention that we have also used the HL2 model to calculate $g(r)$.
HL2 appears less accurate than HL but
better than HL1. We do not plot $g_2(r)$ to avoid obscuring the figures.

%%%%%%%%%%%%%%%%%%%%%%%% Section 6 %%%%%%%%%%%%%%%%%%%%%%%%%%%%%%%%%%%%%
%%%%%%%%%%%%%%%%%%%%%%%  CONCLUSIONS  %%%%%%%%%%%%%%%%%%%%%%%%%%%%%%%%%%
\section{Conclusions}

Thus, the harmonic lattice model allows one
to study static and dynamic properties of quantum and classical
Coulomb crystals. The model is relatively simple, especially
in comparison with numerical methods like MC, PIMC and MD.
The model can be considered as complementary
to the traditional numerical
methods. Moreover, it can be used to explore
dynamic properties of the Coulomb crystals
and quantum effects in the cases where the use of numerical
methods is especially complicated. For instance, the
harmonic lattice model predicts singularities of
the static inelastic structure factor at the positions
of Bragg diffraction peaks.
We expect also that the HL model can describe
accurately non-Coulomb crystals whose lattice
vibration properties are well determined.

{\bf Acknowledgements.}
We are grateful to N.~Ash\-croft for discussions.
The work of DAB and DGY was supported in part by RFBR
(grant 99--02--18099),
% RFBR-DFG (96--02--00177G)
INTAS (96--0542), and KBN (2 P03D 014 13).
The work of HEDW and WLS was performed under the auspices
of the US Dept.\ of Energy under contract number W-7405-ENG-48 for
the Lawrence Livermore National Laboratory and
W-7405-ENG-36 for the Los Alamos National Laboratory.

%%%%%%%%%%%%%%%%%%REFERENCES%%%%%%%%%%%%%%%%%%%%%%%%%%%%%%%%%%%%%%

\end{document}